\documentclass[twocolumn,aps,floatfix,showpacs,showkeys,prl,superscriptaddress]{revtex4}
\pdfoutput=1
\usepackage{epsfig}
\usepackage{graphicx}
\usepackage{amsmath}
\usepackage{amssymb}
\usepackage{amsfonts}
\usepackage{tabularx}
\usepackage{color}
\usepackage{dcolumn}
\usepackage{bm}

\begin{document}

\title{Quantitative magnetic information from reciprocal space maps in transmission electron microscopy}

\author{Hans Lidbaum}
\affiliation{Department of Engineering Sciences, Uppsala University, Box 534, S-751 21 Uppsala, Sweden}
\author{J\'{a}n Rusz}
\affiliation{Department of Physics and Materials Science, Uppsala University, Box 530, S-751 21 Uppsala, Sweden}
\affiliation{Institute of Physics, ASCR, Na Slovance 2, CZ-182 21 Prague, Czech Republic}
\author{Andreas Liebig}
\altaffiliation[Currently at: ]{Institute of Physics, Chemnitz University of Technology, D-09107 Chemnitz, Germany.}
\affiliation{Department of Physics and Materials Science, Uppsala University, Box 530, S-751 21 Uppsala, Sweden}
\author{Bj\"{o}rgvin Hj\"{o}rvarsson}
\author{Peter M. Oppeneer}
\affiliation{Department of Physics and Materials Science, Uppsala University, Box 530, S-751 21 Uppsala, Sweden}
\author{Ernesto Coronel}
\affiliation{Department of Engineering Sciences, Uppsala University, Box 534, S-751 21 Uppsala, Sweden}
\author{Olle Eriksson}
\affiliation{Department of Physics and Materials Science, Uppsala University, Box 530, S-751 21 Uppsala, Sweden}
\author{Klaus Leifer}
\affiliation{Department of Engineering Sciences, Uppsala University, Box 534, S-751 21 Uppsala, Sweden}

\date{\today}

\begin{abstract}
One of the most challenging issues in the characterization of magnetic materials is to obtain quantitative analysis on the nanometer scale. Here we describe how electron magnetic circular dichroism (EMCD) measurements using the transmission electron microscope (TEM) can be used for that purpose, utilizing reciprocal space maps. Applying the EMCD sum rules, an orbital to spin moment ratio of $m_L/m_S=0.08 \pm 0.01$ is obtained for Fe, which is consistent with the commonly accepted value. Hence, we establish EMCD as a quantitative element specific technique for magnetic studies, using a widely available instrument with superior spatial resolution.
\end{abstract}


\maketitle

Fast advances in the field of magnetic nanostructures, both in fundamental research and
technological development, call for new magnetic characterization methods. Electron
microscopy is nowadays a standard technique for structural and chemical analysis down to
the atomic scale. Magnetic imaging \cite{hubert} in the TEM is
also possible while measurements of element-specific magnetic moments have until now
been the domain of synchrotron based dichroic experiments, such as x-ray magnetic
circular dichroism (XMCD) \cite{stohr}. Although XMCD is widely applied in materials science, it
is mainly related to surface measurements and with limitations in spatial resolution. The
work of Schattschneider et al. \cite{nature} -- reporting an observation of dichroic
effects in the TEM -- opened a new route for high-resolution element specific magnetic
characterization, using widely accessible standard laboratory equipment. 

\begin{figure}
  \includegraphics[width=6.5cm]{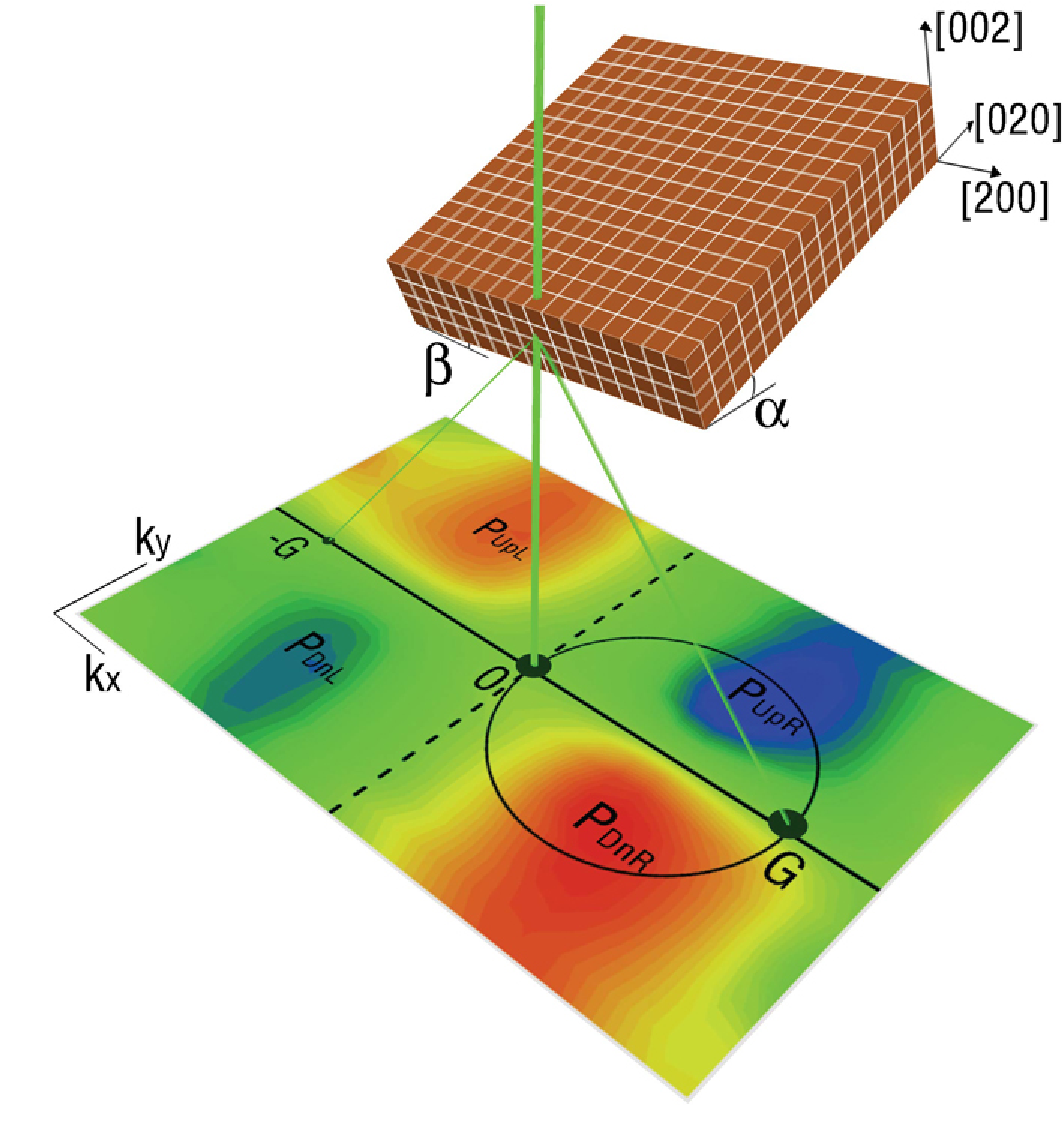}
  \caption{A sketch of the diffraction geometry and distribution of EMCD signal in the $k_x-k_y$ plane.
The sample is tilted with respect to the incoming electron beam from a high symmetry orientation
to the 2BC geometry where $\alpha \approx 10^\circ$ and $\beta \approx 0.4^\circ$. In this geometry only the transmitted (0) and
Bragg scattered beams ($\mathbf{G}$) are strongly excited (large black spots). The optimum detector
positions are indicated as $P_{UpR}$ and $P_{DnR}$ for 2BC geometry using a horizontal mirror axis (solid
line). For the 3BC geometry, $\alpha \approx 10^\circ$ and $\beta = 0^\circ$, an additional vertical mirror axis is available (dashed line). In the
experiment, energy filtered diffraction patterns appear in the $k_x-k_y$ plane; here an EMCD map is
shown as a guide to the eye. The black circle indicates detector positions where $\mathbf{q} \perp \mathbf{q'}$, i.e., the
Thales circle.\label{fig:geometry}}
\end{figure}

EMCD measurements are in principle simple. An unpolarized electron beam, passing
through a magnetic material, exhibits a magnetic dichroism in the momentum resolved
electron energy-loss spectra (EELS) \cite{nature}. The origin of this effect stems from the
inelastic scattering of incoming high-energy electrons that excite core electrons to unoccupied states. The signal at a scattering vector $\mathbf{k}$ contains mixed contributions of all pairs of diffracted beams with momentum transfers $\mathbf{q}$ and $\mathbf{q'}$. A dichroic effect appears
when two EELS-spectra -- extracted at specific detector positions in reciprocal space 
defined by a mirror axis -- are subtracted (see Fig.~\ref{fig:geometry}). This difference spectrum is called in
the following the EMCD signal. 

While the principle of EMCD has been demonstrated, decisive progress is required to
allow quantitative magnetic analysis, which is reported here. The recent
derivation of the EMCD sum rules for extraction of spin ($m_S$) and orbital ($m_L$) magnetic
moments represents an important step in that direction \cite{sumrules,lionelsr}. As EMCD relies on reciprocal
space vectors, proper $\mathbf{k}$-space selection of detector positions is essential. So far, most measurements are carried
out by selecting a limited part of reciprocal space from where the EELS-spectra are
acquired \cite{nature,lacbed}. As shown in this letter, increased flexibility for data optimization and
precision in $\mathbf{k}$-space selection is obtained when a map showing the distribution of the signal
in reciprocal space is available. Energy filtered reciprocal space maps can so far only be
considered as semi-quantitative \cite{lionelsr,lacdif}, since only raw-data was considered. For a quantitative
measurement, the signal to noise (S/N) ratio must be substantially increased and a set of
procedures for data analysis of core-loss edge energy intensities must be devised. 

We choose to demonstrate the technique on bcc Fe, primarily because its magnetic properties
are well known, which allows for a precise exploration of the EMCD technique. Therefore, a Fe sample grown by UHV magnetron sputtering (texture angle of
$\pm 0.3^\circ$) was chosen for the analysis. Three-dimensional data sets, so called data cubes,
consisting of the reciprocal $k_x-k_y$ plane and electron energy-loss are acquired to obtain
maps of the EMCD signal. The data cubes are similar to the spectrum-images introduced by
Jeanguillaume and Colliex \cite{jeang}, but here diffraction patterns instead of images are used. The
energy filtered diffraction patterns were acquired using a Gatan GIF2002
spectrometer on a FEI Tecnai F-30ST FEG operated at 300kV with energy step of 1eV (slit width 2eV) at a sample region of thickness $19 \pm 2$nm and an electron beam diameter of 600nm. 

The \emph{geometrical conditions} of the EMCD experiment are crucial for understanding the effect. The first EMCD experiments \cite{nature} were performed with the
sample oriented in a two beam case (2BC) geometry, see Fig.~\ref{fig:geometry}. The
sample is tilted with respect to the incoming electron beam from a high symmetry
orientation by an angle $\alpha \sim 10^\circ$, where mainly a row of reflections in reciprocal space is excited. By tilting the sample further in the perpendicular
direction by a small angle of $\beta \sim 0.4^\circ$ the 2BC geometry is obtained. In this geometry the
transmitted and Bragg scattered beam $\mathbf{G} = (200)$ in Fe are strongly excited, while all others
are weak. To use the sum rules \cite{sumrules}, allowing a quantitative assessment of the
$m_L/m_S$ ratio, the dichroic signal must be extracted using \emph{two symmetrically placed detectors}
in reciprocal space and the incoming beam should lie within a mirror
symmetry plane of the crystal structure. In 2BC geometry the EMCD signal is given by the difference of a spectrum extracted in the upper
half plane (at $P_{UpR}$) and the corresponding spectrum in the lower half plane (at $P_{DnR}$) in
reciprocal space, see Fig.~\ref{fig:geometry}. This defines a mirror axis for the 2BC
geometry, denoted as \emph{horizontal mirror axis}. By using this mirror axis and correspondingly
subtracting all spectra in reciprocal space, maps of the EMCD signal are
constructed. However, the 2BC geometry does not fulfill the condition of symmetric
detector positions. This can be seen in Fig.~\ref{fig:geometry}, where the (200) atomic planes are not
symmetric with respect to the positions of the detectors ($P_{UpR}$ and $P_{DnR}$).  Then the cancellation of non-magnetic contributions to the signal is not perfect \cite{jmic,theory}, leading to
limitations in the use of the 2BC geometry for $m_L/m_S$ ratio determination. Therefore, we
apply the three beam case (3BC) geometry, which is fully
symmetric. Here, the incoming beam is oriented within a symmetry plane (200), $\beta = 0^\circ$ in
Fig.~\ref{fig:geometry}. The condition of symmetric detector positions is now fulfilled when extracting the
dichroic signal as a difference between left and right half plane, defining a
\emph{vertical mirror axis}. The 3BC geometry also enables the use of a
horizontal mirror axis. Using both mirror axes improves the S/N ratio by using data from all four quadrants. Moreover, it can be shown that such a \emph{double difference} procedure corrects slight misorientations from exact 3BC geometry \cite{theory}.

\begin{figure}
  \includegraphics[width=3.8cm]{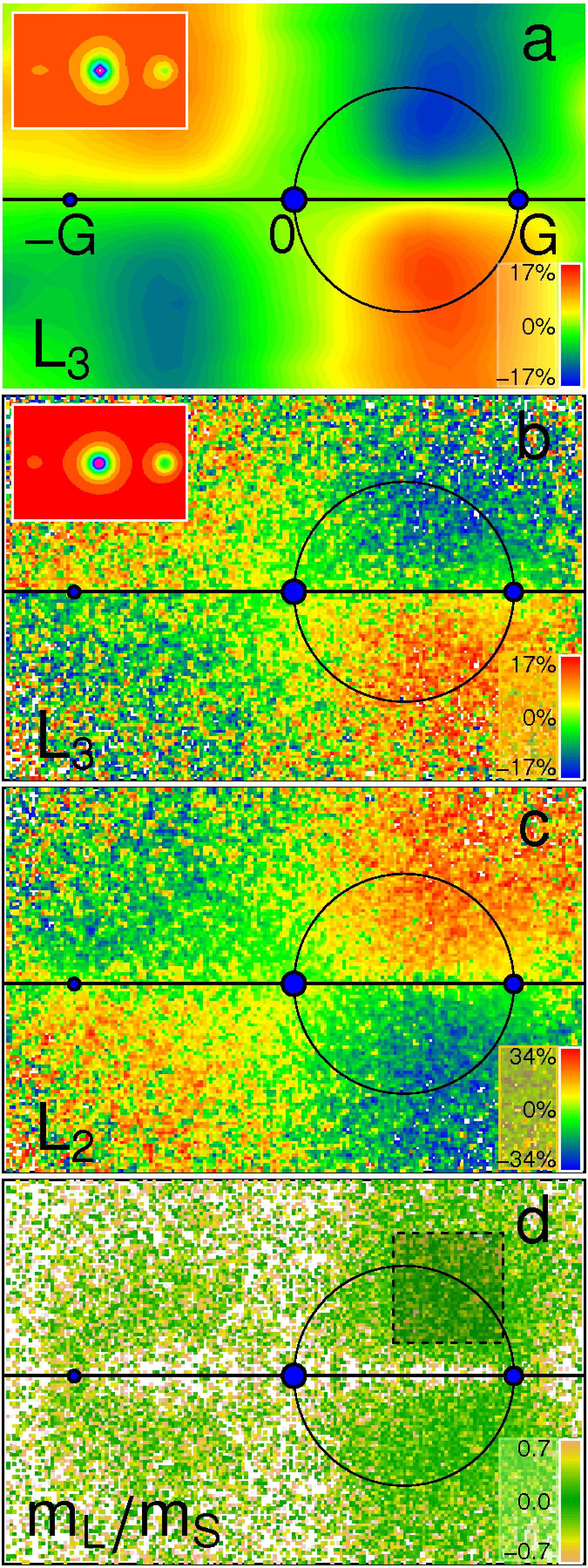}
  \includegraphics[width=3.8cm]{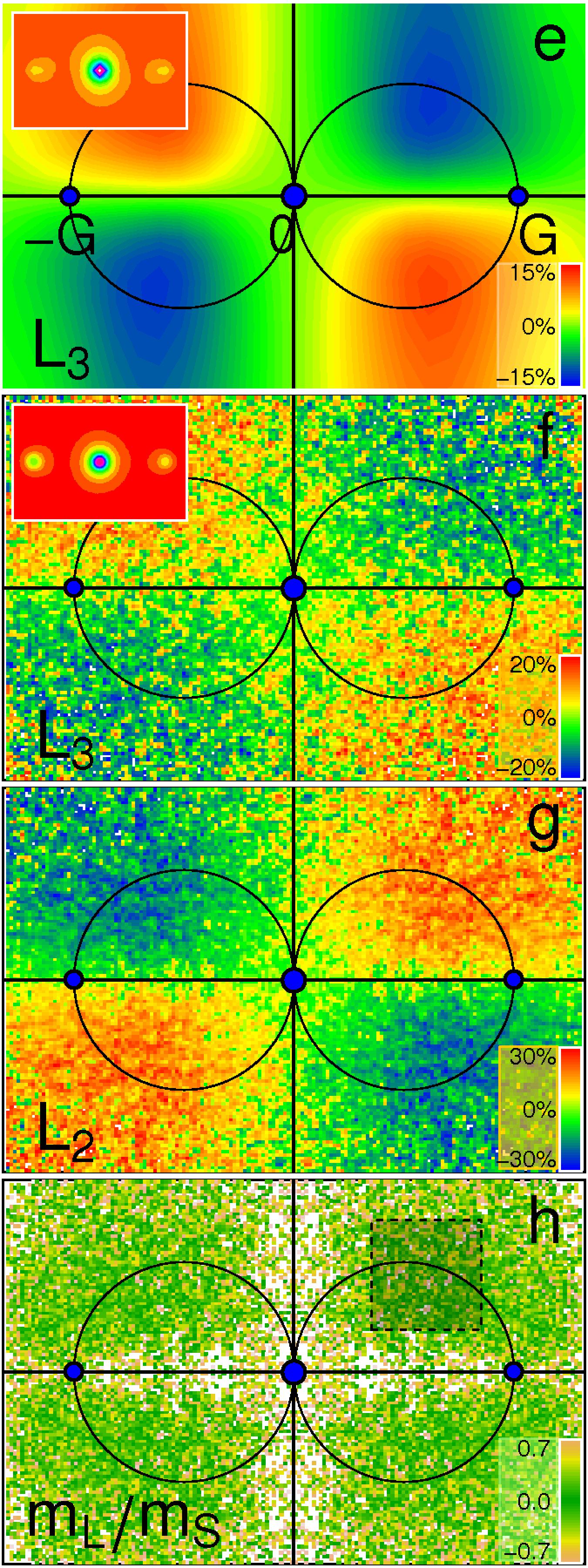}
  \caption{Reciprocal space maps of the EMCD signal and $m_L/m_S$ ratio for the Fe sample oriented in a
2BC geometry (left) and 3BC geometry (right). Theoretical relative EMCD maps at the $L_3$ edge are shown in a) and e). The inset shows the simulated diffraction pattern. In b) and c) [f) and g) for 3BC] maps of experimentally obtained relative EMCD signal at $L_3$ and $L_2$ edges are shown. The black lines indicate the applied mirror axes and blue spots the positions of the transmitted and Bragg scattered $\mathbf{G} = (200)$ and $-\mathbf{G} = (\bar{2}00)$ beams. The insets in b) and f) show the diffraction patterns averaged over an energy interval from 695 eV to 740 eV. In d) and h) the experimental $m_L/m_S$ maps are shown, where the box indicates the region with minimal noise. Values outside the range of color bar are shown as white. \label{fig:2bc3bc}}
\end{figure}

The EMCD signal evolves as an entwined property of sample thickness and orientation due
to dynamic electron diffraction effects. Thus, simulations of the EMCD experiments
include both Bloch-wave simulations of dynamic scattering of the fast electrons, as well as
\emph{ab initio} calculations of the inelastic scattering in a magnetic sample \cite{papertheory}. Inelastic
transition matrix elements were calculated including spin-orbit coupling with magnetization
direction [016] as imposed by the magnetic field of the objective lens, corresponding to
$\alpha \approx10^\circ$. In order to simulate the finite collection angle and the relative
misorientation of illuminated regions in the experiments ($\pm 0.3^\circ$), maps of the EMCD signal
were averaged over a set of Laue circle centers. Simulated maps of the relative EMCD
signal at $L_3$ edge for 2BC and 3BC geometries are displayed in Fig.~\ref{fig:2bc3bc}a and \ref{fig:2bc3bc}e,
respectively. Since DFT calculations predict a very low orbital momentum ($\sim 0.045 \mu_B$) in
bcc Fe, the EMCD map at $L_2$ has a similar structure as the $L_3$
map but with reversed sign \cite{papertheory}. In the 2BC geometry, the strongest dichroic signal is shifted towards the strongly excited reflection, denoted as $\mathbf{G}$ in Fig.~\ref{fig:geometry}, in agreement with Refs. \cite{JAP,verbeeck}. A reduced EMCD signal is also present around the weakly excited reflection ($-\mathbf{G}$).

\begin{figure}
  \includegraphics[width=7.8cm]{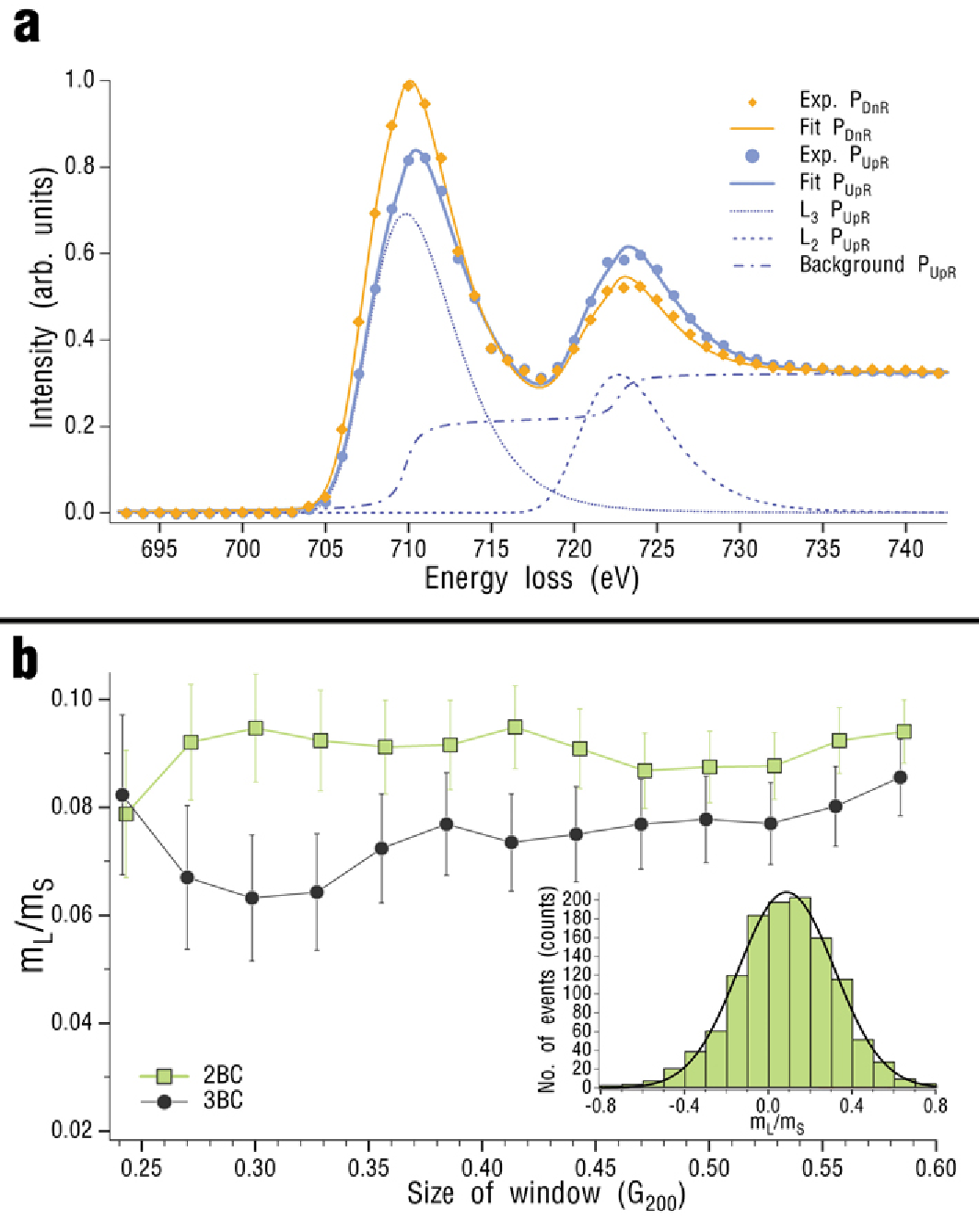}
  \caption{Experimental EELS-spectra at Fe $L_{2,3}$ edges and $m_L/m_S$ ratios as a function of collection
window size in $m_L/m_S$ ratio maps. The two spectra in a) were extracted from the experimental
data cube with the sample in 2BC geometry at the $P_{DnR}$ (orange squares) and $P_{UpR}$ positions
(blue circles), averaging over $0.5G_{200} \times 0.5G_{200}$ in reciprocal space. The components of the fit function for the spectrum at the $P_{UpR}$ position are shown. 
b) $m_L/m_S$ ratio obtained for various window sizes in 2BC geometry
using horizontal mirror axis and in 3BC using double difference method. Inset shows the histogram with fit of $m_L/m_S$ ratio for the window size of $0.5G_{200} \times 0.5G_{200}$ in 2BC orientation (indicated in Fig.~\ref{fig:2bc3bc}d).
\label{fig:lsratio}}
\end{figure}

After applying cross-correlation \cite{CC} on the transmitted beam, each spectrum in the experimental data-cube was pre-edge background subtracted (power-law model) and normalized in the post-edge region (at 736-738eV as seen in Fig.~\ref{fig:lsratio}a). After that, each spectrum in the $k_x-k_y$ plane was peak fitted \cite{peakfit} and the integrated $L_{2,3}$ edge intensities are used to construct the EMCD maps. 
Peak fitting removes the need of somewhat arbitrary energy window selection for separating $L_3$ and $L_2$ components and corrects for eventual energy shifts due to non-isochromaticity of the spectrometer.
Experimental maps of the relative EMCD signal at the $L_{2,3}$ edges in
2BC and 3BC geometries are shown in panels b, c, f and g of Fig.~\ref{fig:2bc3bc}, respectively. We find a very good correspondence between the theoretical prediction (panels a and e) and measurements regarding the position in reciprocal space and strength of the EMCD signal for both geometries. From the 2BC data cube two spectra were extracted using a box with the size of $0.5\mathbf{G}_{200} \times 0.5\mathbf{G}_{200}$ at the $P_{UpR}$ and $P_{DnR}$-positions (similarly to Ref.~\cite{verbeeck}) and peak fitted. 
In Fig.~\ref{fig:lsratio}a the $L_{2,3}$ edge intensities and a continuum background
used in the fit of the experimental data are shown. The spectra show a clear EMCD signal
with good S/N ratio. Experiment and simulation reveal a relative EMCD signal (difference
divided by its sum) at the $L_3$ edge of approx. 8\% and 12\%, respectively. Note that a strong
EMCD signal is obtained in spite of the texture angle of $\pm 0.3^\circ=\pm 5$mrad (corresponding to $\pm 0.4G_{200}$), which is of high importance for a practical use of EMCD. Thus the technique is robust against small distortions of the crystal orientation, allowing for the characterization of imperfect crystals. A texture angle will have a similar effect on the EMCD signal as an increase of convergence angle since both can be viewed as electrons hitting a single crystalline lattice at slightly different angles. A semi-convergence angle of 10 mrad corresponds in the TEM used here to a probe size of about 0.13 nm \cite{klenov}. In this work, we show that a quantitative EMCD signal is obtainable at deviations in the relative sample to probe orientation of $\pm 5$mrad. Therefore we believe that the approach demonstrated here has the potential for sub-nanometer resolution. This argumentation is underlined by 
the recent demonstration of an EMCD signal using a convergent electron beam at 2nm resolution \cite{2nm}. The quest to reach the ultimate resolution of this technique will thus not be dominated by the achievable convergence angle or electron probe size, but by the stability of the sample and the electron microscope itself.

A quantitative assessment of spin and orbital angular momenta can be obtained from the
EMCD sum rules \cite{sumrules,lionelsr}. Correspondingly, a map of the $m_L/m_S$ ratio is obtained by
calculating at each position in reciprocal space:
\begin{equation} \label{eq:ls}
  \frac{m_L}{m_S} = \frac{2}{3} \frac{\int_{L_3} \Delta I(E) dE + \int_{L_2} \Delta I(E) dE}{\int_{L_3} \Delta I(E) dE - 2\int_{L_2} \Delta I(E) dE}
\end{equation}
with $\Delta I(E) = I(P_i) - I(P_j)$ 
where $I(P_i)$ and $I(P_j)$ are the EELS-spectra extracted at the two detector positions, e.g. in
the 2BC geometry a choice of $P_i = P_{UpR}$ implies using $P_j = P_{DnR}$ see Fig.~\ref{fig:geometry}. The
general sum rule expressions \cite{sumrules} predict a $\mathbf{k}$-independent $m_L/m_S$ ratio, i.e. the
intensity of a large part of reciprocal space can be employed to determine the ratio.
Nevertheless, the S/N ratio should be optimized by selecting the region with highest EMCD signal. 
Since sum rules involve several approximations and assumptions, which may not be perfectly fulfilled, we evaluated theoretical $m_L/m_S$ maps (not shown). They reveal the effect of 2BC asymmetry \cite{jmic}, which becomes non-negligible in regions outside the Thales circle. In the 3BC geometry even a deviation as small as $\beta=0.05\mathbf{G}_{200}$ can introduce substantial variations of the observed $m_L/m_S$ ratio throughout the diffraction plane. Detailed calculations have shown that construction of the double difference map, i.e., exploiting \emph{both} mirror axes, is a very efficient method for correcting these inaccuracies \cite{theory}.

The maps enable an accurate determination of the $m_L/m_S$ ratio. A selection window is used
in these maps from which the values of all individual pixels are extracted,
thus optimizing the S/N ratio. The overall $m_L/m_S$ ratio for a given window size (collection
angles) is obtained by fitting the histogram of the individual $m_L/m_S$ values with a Gaussian
(see inset in Fig.~\ref{fig:lsratio}b). The ratio is determined as the center of the Gaussian fit. Plotting this ratio as a function of the window size, a close to constant value in the $m_L/m_S$ ratio is
reached when the size of the box is sufficient to reduce the statistical fluctuations. If the
window gets too large, regions with low S/N ratio are included which increase the
standard deviation of the $m_L/m_S$ ratio. Regions of minimal noise in the experimental $m_L/m_S$
maps (indicated in panels d and h of Fig.~\ref{fig:2bc3bc}) correlate with local maxima found in the EMCD maps. 
These results agree well with findings in Ref.~\cite{verbeeck}, although here we apply the
integration window numerically on the $m_L/m_S$ ratio rather than on maps of the EMCD signal.
This provides better flexibility for numerical analysis and a reliable error bar.
Here,
an optimum size of the window in reciprocal space is typically $0.5G_{200} \times 0.5G_{200}$. We
obtain a consistent $m_L/m_S$ ratio, depicted in Fig.~\ref{fig:lsratio}b, of $0.09 \pm 0.01$ in the 2BC and $0.08 \pm 0.01$
in 3BC geometry using the double difference maps. The standard error $s = 0.01$ was
estimated using $N=1225$ individual $m_L/m_S$ ratios within the selection window (standard deviation of an individual $m_L/m_S$ ratio is $\sigma$=0.23, $s$=$\sigma$/$\sqrt{N}$). 

Our experimental $m_L/m_S$ ratio for bcc Fe is $0.08 \pm 0.01$.
The most commonly accepted values of the
ratio are obtained by XMCD and gyromagnetic ratio measurements, 0.043 \cite{chen} and 0.044
\cite{reck}, respectively. One may observe that the present value is somewhat larger. However, neutron scattering experiments give a value closer
to the presently obtained value, namely 0.062 \cite{landolt}. For bcc Fe the orbital moment is small
and it is not surprising that this delicate property shows certain dispersion.
E.g., with XMCD, values of 0.043 \cite{chen}, 0.07 \cite{obrien}, 0.085 \cite{arvanitis} and
$0.12 \pm 0.05$ \cite{zaharko} for the $m_L/m_S$ ratio have been reported for bcc Fe. In the experiments
presented here, oxygen is most likely present in the vicinity of one of the Fe surfaces,
which may also influence the $m_L/m_S$ ratio. 
It should be noted that one previous EMCD experiment measuring the 
$m_L/m_S$ ratio has been conducted for Fe \cite{lionelsr}. Although the $m_L/m_S$ ratio reported in that work is close to ours, their analysis lacked a quantitative data treatment (e.g., peak fitting), as they stated explicitly.

Experimental and theoretical maps of the EMCD signal in reciprocal space are shown to
agree well for the investigated sample geometries. 
Asymmetry problems are successfully compensated by the double difference method within the 3BC geometry.
Obtained $m_L/m_S$ ratios of bcc Fe are close to the commonly accepted value. Our work establishes therefore the quantitative use of EMCD, which, together with the high spatial resolution available in modern TEMs promises unique possibilities for truly nano-scale magnetic studies.

\begin{acknowledgments}
The work was supported through the Swedish Research Council (VR), Knut and Alice
Wallenberg foundation (KAW), G\"{o}ran Gustafsson foundation, STINT and computer cluster \textsc{david} of IOP ASCR
(AVOZ10100520).
\end{acknowledgments}


\begin{thebibliography}{99}

\bibitem{hubert} A. Hubert, R. Sch\"{a}fer, Magnetic Domains - The analysis of magnetic microstructures, Springer, Berlin, 1998.

\bibitem{stohr} J. St\"{o}hr et al., Science \textbf{259}, 658 (1993).

\bibitem{nature} P. Schattschneider et al., Nature {\bf 441}, 486 (2006).

\bibitem{sumrules} J. Rusz et al., Phys. Rev. B {\bf 76}, 060408(R) (2007).

\bibitem{lionelsr} L. Calmels et al., Phys. Rev. B {\bf 76}, 060409(R) (2007).

\bibitem{lacbed} P. Schattschneider et al., Ultramic. \textbf{108}, 433 (2008).

\bibitem{lacdif} B. Warot-Fonrose et al., Ultramic. \textbf{108}, 393 (2008).

\bibitem{jeang} C. Jeanguillaume, C. Colliex, Ultramic. {\bf 28}, 1 (1989).

\bibitem{jmic} J. Rusz et al., J. Microsc., accepted.

\bibitem{theory} J. Rusz, submitted to Phys. Rev. B (BW10795).

\bibitem{papertheory} J. Rusz, S. Rubino, P. Schattschneider, Phys. Rev. B {\bf 75}, 214425 (2007).

\bibitem{JAP} P. Schattschneider et al., J. Appl. Phys. {\bf 103}, 07D931 (2008).

\bibitem{verbeeck} J. Verbeeck et al., Ultramic. {\bf 108}, 865 (2008).

\bibitem{CC} B. Schaffer et al., Ultramic. {\bf 102}, 27 (2004).

\bibitem{peakfit} F. Wang et al., Micron {\bf 37}, 316 (2006); R. Hesse et al., Surf. Interface Anal. {\bf 39}, 381 (2007).

\bibitem{klenov}  D. O. Klenov et al., Phys. Rev. B {\bf 76}, 014111 (2007).

\bibitem{2nm} P. Schattschneider et al., Phys. Rev. B {\bf 78}, 104413 (2008).

\bibitem{chen} C.~T. Chen et al., Phys. Rev. Lett. {\bf 75}, 152 (1995).

\bibitem{reck} R. A. Reck, D. L. Fry, Phys. Rev. {\bf 184}, 492 (1969).

\bibitem{landolt} M. B. Stearns, Landolt-B\"{o}rnstein Numerical Data and Functional Relationships in Science and
Technology, (Springer, Berlin, 1986), Group 3, Vol. 19, Pt. a , p.53.

\bibitem{obrien} W. L. O'Brien et al., J. Appl. Phys. {\bf 76}, 6462 (1994).

\bibitem{arvanitis} D. Arvanitis et al., in Lecture Notes in Physics (Springer, Berlin/Heidelberg, 1996), vol. 466, pp. 145-157.

\bibitem{zaharko} O. Zaharko et al., Eur. Phys. J. B {\bf 23}, 441 (2001).

\end{thebibliography}
\end{document}